# Spurious magnetism in high-$T_c$ superconductor

P.K. Mang[1], S. Larochelle[2], and M. Greven[1,3]

Departments of [1]Applied Physics, [2]Physics, and [3]Stanford Synchrotron Radiation Laboratory,
Stanford University, Stanford, CA 94305, USA

One challenge in condensed-matter physics is to unravel the interplay between magnetism and superconductivity in copper oxides with a high critical temperature ($T_c$). Kang et al.[1] claim to have revealed a quantum phase transition from the superconducting to an antiferromagnetic state in the electron-doped material $Nd_{2-x}Ce_xCuO_4$ (NCCO) based on the observation of magnetic-field-induced neutron scattering intensity at (1/2,1/2,0), (1/2,0,0), and related reflections. Here we argue that the observed magnetic intensity is due to a secondary phase of $(Nd,Ce)_2O_3$. We therefore contend that the effect is spurious and not intrinsic to superconducting NCCO.

To achieve superconductivity in NCCO, a rather severe oxygen-reduction procedure has to be applied[2]. We have discovered that the reduction process decomposes a small (0.01% –0.10%) volume fraction of NCCO. The resultant $(Nd,Ce)_2O_3$ secondary phase has the complex cubic bixbyite structure, common among rare-earth (RE) sesquioxides[3], with a lattice constant, $a_c$, that is about $2\sqrt{2}$ times the planar lattice constant of tetragonal NCCO. $(Nd,Ce)_2O_3$ is epitaxial with the host lattice, with long-range order parallel to the $CuO_2$ planes of NCCO, but extending only about $5a_c$ perpendicular to the planes. Because of the relationship between the two lattice constants, certain structural reflections from the impurity phase appear at seemingly commensurate NCCO positions — that is, the cubic $(2,0,0)_c$ reflection can also be indexed as (1/2,1/2,0). However, there is approximately a 10% mismatch between $a_c$ and the c-lattice constant of NCCO, and therefore $(0,0,2)_c$ can also be indexed as (0,0,2.2).

There are 32 rare-earth ions in the $RE_2O_3$ unit cell, belonging to two crystallographically distinct sites with inequivalent saturated moments[3]. At the $(2,0,0)_c$ reflection, the contributions from the two rare-earth sites interfere destructively, which should lead to a peak in the observed scattering intensity in the paramagnetic phase if the moments saturate at different fields. Although the magnetic structure and spin hamiltonian of epitaxial, quasi-two-dimensional $(Nd,Ce)_2O_3$ are unknown, it is possible to devise simple experiments to test whether the field-induced scattering is due to NCCO or $(Nd,Ce)_2O_3$.

Kang et al. find that at a temperature of 5 K, the (1/2,1/2,0) (that is, $(2,0,0)_c$) intensity reaches a peak at a field of about 6.5 T, and argue that this peak is associated with the upper critical field $B_{c2}$ of NCCO. Figure 1a summarizes the field dependence of an x=0.18 superconducting sample of ours in the temperature range 1.9–10 K. Our data agree with those of Kang *et al*. The figure demonstrates that the intensity scales with B/T and exhibits a peak consistent with two-moment paramagnetism. Furthermore, as the upper critical field of a superconductor increases with decreasing temperature, this implies that the reported correspondence of the peak position with $B_{c2}$ at 5 K is coincidental. We do not observe spontaneous neodymium ordering of either $(Nd,Ce)_2O_3$ or NCCO down to 1.4 K.

Figure 1b,c shows that the field effects reported by Kang *et al*. are also observable in a non-superconducting, oxygen-reduced, x=0.10 sample, both at the previously reported positions and at positions that are unrelated to the NCCO lattice but equivalent in the cubic lattice of $(Nd,Ce)_2O_3$. Not only are the incommensurate positions (0,0,2.2) and (1/4,1/4,1.1) unrelated to the proposed NCCO magnetic order, but the physical situation of the magnetic field applied parallel (in the cases of the (0,0,2.2) and (1/4,1/4,1.1)) or perpendicular (in all other cases) to the $CuO_2$ planes is fundamentally different in that the upper critical fields for the two geometries differ significantly. Note that (1/2,0,0) and (1/4,1/4,1.1) correspond to $(1,1,0)_c$ and $(1,0,1)_c$, respectively. Care was taken to ensure that in all cases the magnetic field was applied along a cubic axis of $(Nd,Ce)_2O_3$ and perpendicular to the scattering wavevector.

These simple experimental tests demonstrate that the observed field effects in oxygen-reduced NCCO result from an epitaxial secondary phase of $(Nd,Ce)_2O_3$.

1. Kang, H.J. *et al*. *Nature* **423**, 522-525 (2003).
2. Tokura, Y., Takagi, H. & Uchida, S. *Nature* **337**, 345-347 (1989).
3. Moon, R.M., Koehler, W.C., Child, H.R. & Raubenheimer, L.J. *Phys. Rev.* **176**, 722-731 (1968).

**Figure 1** Field and temperature dependence of magnetic scattering. **a**, Arbitrarily scaled scattering intensity at (1/2,1/2,0) for a superconducting sample of NCCO (nominal cerium concentration *x*=0.18; $T_c$=20K) as a function of *B/T* with the field along [0,0,1]. The results are compared with the data of Kang *et al.*[1] (*x*=0.15; T=5 K). **b,c**, Comparison of the results of Kang *et al*. with data taken at *T*=4K for a superconducting sample (*x*=0.18) and a non-superconducting sample (*x*=0.10). Superconductivity in NCCO can be achieved only for *x*>0.13. The magnetic field is applied along [1,-1,0] for (0,0,2.2) and (1/4,1/4,1.1) and along [0,0,1] in all other cases. Data were normalized by maximum intensity. Full details are available from the authors. Structural sample characterization was carried out at the Stanford Synchrotron Radiation Laboratory, while magnetic neutron scattering measurements were performed at the NIST Center for Neutron Research and the Berlin Neutron Scattering Center.

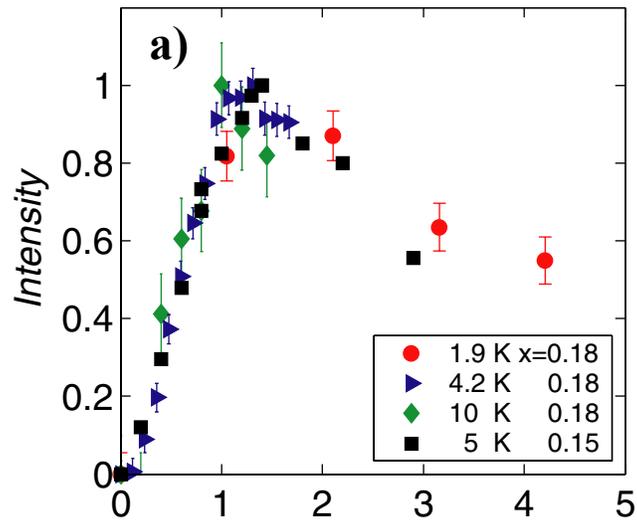
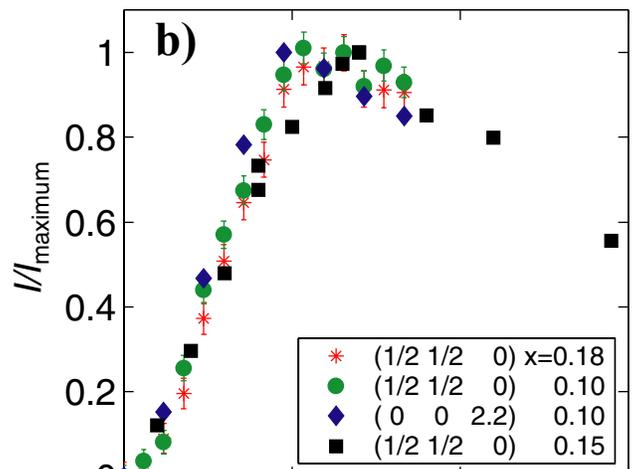
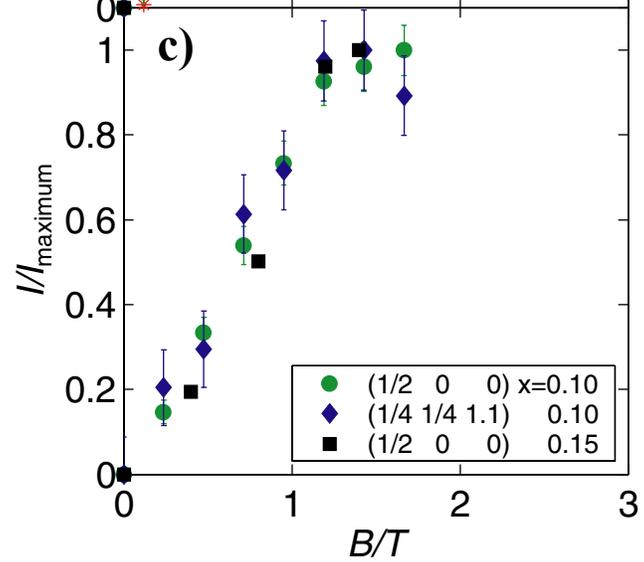